\documentstyle[12pt]{article} 

\newcommand{\bref}[1]{(\ref{#1})} 

\newcommand{\be}{\begin{equation}} 
\newcommand{\ee}{\end{equation}} 

\def\lsim{\mathrel{\rlap{\lower4pt\hbox{\hskip1pt$\sim$}}
    \raise1pt\hbox{$<$}}}
\def\gsim{\mathrel{\rlap{\lower4pt\hbox{\hskip1pt$\sim$}}
    \raise1pt\hbox{$>$}}}
\def\frac#1#2{{{#1} \over{#2}}} 

\begin{document}  
\begin{titlepage} 
  
\begin{flushright}  
{LBNL-47547 \\}
{\hfill November 2001 \\ }  
\end{flushright}  
\vglue 0.2cm  
	   
\begin{center}   
{ 
{Characterization of a Branch of the Phylogenetic Tree \\ }  
\vglue 1.0cm  
{Stuart A. Samuel \\ }   
\vglue 0.5cm  

{\it Lawrence Berkeley Laboratory\\}
{\it MS 50A-5105\\}
{\it Berkeley, CA 94720 USA\\}  

\vglue 1.2cm  

{Gezhi Weng \\ }   
\vglue 0.5cm  

{\it Department of Pharmacology\\}
{\it Mount Sinai School of Medicine\\}
{\it New York, NY 10029, USA\\}

\vglue 0.8cm

{\bf Abstract}  
}  
\end{center}   
{ \rightskip=3pc\leftskip=3pc  
\quad We use a combination of  
analytic models and computer simulations to gain insight  
into the dynamics of evolution.  
Our results suggest that certain interesting phenomena  
should eventually emerge from the fossil record. 
For example, 
there should be a ``tortoise and hare effect'': 
Those genera with the smallest species death rate 
are likely to survive much longer than genera with 
large species birth and death rates. 
A complete characterization of the behavior  
of a branch of the phylogenetic tree corresponding to a genus
and accurate mathematical representations  
of the various stages are obtained. 
We apply our results to address certain controversial issues  
that have arisen in paleontology 
such as the importance of punctuated equilibrium 
and whether unique Cambrian phyla have survived 
to the present.   
} 
\vfill
\eject 
\end{titlepage}

\newpage  

\baselineskip=20pt  

{\bf\large\noindent I.\ Introduction}
\vglue 0.2cm

The fossil record is incomplete,  
providing only a fraction of the species that actually existed. 
This makes it difficult to characterize the phylogenetic tree.
Indeed, different behaviors 
have been suggested (Gould et al., 1977; Gould, 1989), 
and phenomena that might be present are easily overlooked.  
In addition, controversial issues such as 
the importance of punctuated
equilibrium (Eldredge \& Gould, 1972; Stanley, 1979; Gould \& Eldredge, 1993) 
and whether
Cambrian phyla extend to the present (Gould, 1989; Morris, 1998)  
remain unresolved. 
In this work, we bypass the difficulty 
of fossil-record incompleteness 
by using computer simulations and analytic methods 
that lead to a new understanding of the dynamics of evolution.  
We find that the development of a branch 
corresponding to a genus of species 
is characterized 
by three stages with distinct behaviors. 
We establish new types of contingency, 
prove that all genera must eventually go extinct, 
and uncover some surprising phenomena that should be 
present if the fossil record were known more precisely.  

For the most part, 
our analysis is for the evolution of a set of closely related species 
who orginate from a single lineage. 
We shall refer to this as a genus, 
a genus being a group of species with similar 
characteristics usually determined subjectively by taxonomists 
on the basis of morphology. 
One can imagine sudden and gradual origins for a new genus:
An individual through faulty genetic reproduction 
may be radically different from others  
and through subsequent mating may eventually produce 
a significantly different species. 
Alternatively, as a group, 
a species may evolve over a long period of time 
to become sufficiently morphologically different 
from other members in its genus to warrant 
the establishment of a new genus. 
Independent of the creation mechanism, 
it is expected that a new genus begins with a single species. 
Such a species through splitting processes can yield a genus 
with two or more members with similar characteristics. 

Our results are mostly applicable 
to the case of genera as opposed to higher taxonomic groupings 
such as families and orders. 
The reason for this is that the model we use 
assumes that all member species  
have the the same speciation birth
and death rates. 
This is expected to be the case for genera 
but less so for higher groupings. 
Nevertheless, 
some of the phenomena that we have observed for genera 
can be expected to arise for certain families, orders, 
classes and phyla.
It is also possible to generalize the model to include 
larger parts of the phylogenetic tree but that goes 
beyond the aims of the current research. 

\medskip

{\bf\large\noindent II.\ The One-Genus Model}
\vglue 0.2cm

Consider the emergence of a new genus ${\cal G}$ 
initially containing only one species. 
Through genetic mutations, environmental factors and other effects, 
new species arise and old species die 
to create or destroy a lineage of the phylogenetic tree. 
Let $B$, the species birth rate, 
be the probability per unit time that 
a new species line splits off of an old species line. 
Likewise, let $D$, the species death rate, 
be the probability per unit time that 
a species goes extinct. 
It is assumed that $B>D$; 
otherwise the genus soon goes extinct. 
In a small time interval $\Delta t$, 
$d = D \Delta t$, $b = B \Delta t$ 
and $u=1 - d - b$ are respectively 
the probabilities that a particular species dies, 
that it gives rise to an additional lineage and 
that neither of these happens. 
If $N(t)$ is the number of species in the genus, then 
\be
     N(t + \Delta t) = N(t) + \delta N(t)
\quad , 
\label{eq1}
\ee
where $\delta N(t)$ is determined 
by the probabilities $d$, $b$ and $u$.  
This is a stochastic process. 

Since Earth's ecosystem cannot support 
an unlimited number of living organisms, 
we imposed the constraint that $N (t) \le N_{max}$ 
for some maximum number of species $N_{max}$. 
The parameters $B$, $D$ and $N_{max}$ 
vary significantly from genus to genus, 
with $B$ and $D$ generally being larger for smaller organisms. 
For most metazoa, $B$ and $D$ 
range roughly from one species per 10,000 years 
to one per 10 million years (May, 1995; McCune, 1997).
One expects $N_{max}$ to vary from a small number to hundreds. 
As an example, data exist that imply $N_{max} \approx 16$ 
for the grazing horses in North America 
during the Miocene (MacFadden \& Hulbert, 1986). 
For the most part, 
one expects $B$, $D$ and $N_{max}$ to be slowly
varying functions of time. 
In our model, 
we take these parameters to be constant 
(See, however, Section V where $N_{max}$ is allowed to undergo 
jumps to simulate the effects of punctuated equilibrium). 
This approximation will not change our conclusions 
as long as the fractional variations in the parameters 
are significanly less than the smallest inverse time scale in the model, 
that is, 
$1/B dB/dt << D$, $1/D dD/dt << D$ and $1/N_{max} d N_{max}/dt << D$.

The above defines what we mean by the term {\it evosystem}:  
the evolution of a group of species in a constrained environment. 
We wrote a software package to generate and explicitly
display phylogenetic trees 
governed by the probabilities $b$, $d$ and $u$. 
Hundreds of trees were examined to gain insight 
into general behavior 
and interesting phenomena. 
The computer simulations reveal that 
three stages characterize 
the phylogenetic dynamics of a genus 
of the evolutionary tree:  
(1) an initial period of precariousness, 
(2) a period of exponential growth and 
(3) a period of quasi stability.  
See Figure 1a. 

\medskip

{\bf\large\noindent III.\ The Three Stages}
\vglue 0.2cm

During Stage (1), when $N(t)$ is small, 
there is a certain probability $P_{e}$ that a genus goes extinct
because of ``bad luck'' since $B > D$ favors growth. 
See Figure 1b. 
An approximate equation for $P_{e}$ is $P_{e}=D/B$. 
See Appendix A for a derivation. 
This formula becomes exact for $N_{max}=\infty$ 
and is quite accurate for all but small $N_{max}$. If 
the number of species 
is initially $N_0$ instead of 1, 
\be 
     P_{e} = (D/B)^{N_0} 
\quad . 
\label{eq2}
\ee
An analogous result is known for population models (MacArthur, 1972). 
The exact formula when $N_{max} = \infty$ 
for the probability of extinction as a function 
of time is provided in eq.\bref{eq7} below 
and rapidly approaches eq.\bref{eq2} for $t > 1/(B-D)$. 

If ${\cal G}$ survives the period of precariousness, 
Stage (2) arrives. 
It is characterized by exponential growth: 
$\bar N(t) = N_0 \exp (Gt)$ where $G=B-D$ and $\bar N(t)$ 
is the average value of $N(t)$. 
This result is expected and 
is similar to the situation 
in population dynamics (Renshaw, 1991). 
In general, when there is ample vacant phase space for an evosystem, 
species proliferation occurs. 
Such opportunities arise when life moves 
into a new ecological niche, 
when a new morphological development arises, 
or just after a catastrophic extinction. 
The fastest growth in the absolute number of new species $N_{new}$
occurs near the end of Stage (2) since 
${{d \bar N_{new} (t)>} / {dt}} = B \bar N(t) $
and the constraint of $N(t) \le N_{max}$ 
is not operative. 

Stage (3) is characterized by having $N(t)$ near $N_{max}$: 
The genus has expanded fully into its evosystem's phase space. 
The rate of {\it new} species generation 
slows from $B N $ to $D N$ because 
species creation can only occur 
at the expense of the death of a species lineage. 
Thus, the rapid flow of Stage (2) gives way 
to a slower phylogenetic movement. 
See the right side part of Figure 1a. 
During Stage (3), 
adaptation leads to the evolution of old species through natural selection 
but this process is more gradual.  

One would think that a genus could last forever 
in the presence of a stable, rich, life-supporting environment. 
Surprisingly, this is not the case as we demonstrate below. 
It turns out that ${\cal G}$ is guaranteed to go extinct, 
but the time scale $T_e$ for eventual extinction is quite large:  
$T_e = 1/D f_e (B/D)$ with $f_e$ of the order of $ (B/D)^{N_{max}}$.  
The time at which a particular genus 
goes extinct involves considerable uncertainty 
but, since it must, 
Stage (3) is characterized 
as a period of {\it quasi stability}. 
A genus has the best chance to survive 
for a long time if $N_{max}$ is large and $D/B$ is small. 

The result of guaranteed extinction does not take into 
account the following effect. 
One species in the genus may become sufficiently different from the rest 
as to warrant its classification as a new genus. 
In such a situation, it is removed from the genus under consideration 
and becomes the first species of a new genus. 
An example of this is the case of birds and dinosaurs. 
Assuming that "being a reptile" is part of the definition of a dinosaur, 
then the dinosaurs went extinct $65$ million years ago, 
while birds, which descended from the dinosaurs, survived.  
If, on one hand, the new genus occupies the same ecosystem phase space, 
then the $N_{max}$ constaint applies to the combined system: 
$N_{original\ genus} + N_{new\ genus} \le {N_{max}}$.  
Such a model is analyzed in Section VII. 
If, on the other hand, the new genus moves 
into a new region of ecosystem phase space, 
then it will be limited by a new maximum-number-of-species constaint. 
In this case, 
the system resembles two, independent, one-genus models. 
What is more likely to transpire is something between these two extremes. 
Such a system can be modeled using the constraints: 
$N_{o} \le N^o_{max}$, $N_{n} \le N^n_{max}$ and 
$N_{o} + N_{n} \le N^{combined}_{max}$, 
where $N_{o}$ is the number of species in the original genus, 
$N_{n}$ is the number of species in the new genus, 
and $N^o_{max}$, $N^n_{max}$ and $N^{combined}_{max}$ are constants. 

The role of contingency in evolution has been emphasized 
by Gould (1989) and others.  
The extinction effects of Stages (1) or (3) described above  
are specific realizations of contingency 
that are different from those envisioned by Gould 
but are nonetheless present. 

One application of Stage (1) precariousness 
is the ``smearing in time'' of a catastrophic extinction. 
Initial opposition arose to the asteroid impact explanation 
of the KT mass extinction (Alvarez et al., 1980)  
because paleontologists found evidence in sedimentary rock 
that all species did not 
die instantly (Archibald, 1981; Hickey, 1981; Sloan et al., 1986). 
Many researchers assume that the explanation for this is 
an imperfect fossil record. 
This may very well be the reason, 
however, Stage (1) precariousness may be contributing: 
Suppose for example that $B=$1/(200,000 years) 
and $D=$1/(400,000 years) and 
that after the asteroid struck, 
10\% of genera were left with a single species and 
20\% were left with two species. 
Then from eq.\bref{eq2}, 
one concludes that 
10\% of genera would go extinct 
over the next few hundred thousand years. 
In other words, in a catastrophic extinction, 
most genera die out quickly 
but a non negligible fraction die out 
over a much longer period of time. 
Figure 2a shows a computer simulation that exemplifies this 
for a particular genus.

\medskip

{\bf\large\noindent IV.\ The Statistical Mechanics of Evolutionary Trees}
\vglue 0.2cm

Let us perform a statistical analysis. 
Defined $p_N (t)$ to be the probability that  
a tree has $N$ branches at time $t$ on average. 
Each process associated with eq.\bref{eq1}  
produces a ``path'' for $N(t)$,  
a random walk in the interval $0$ to $N_{max}$.  
Averaging over all possible stochastic processes  
produces the probabilities $p_N (t)$.
Differential equations determine the $p_N (t)$: 
$$
{ {d p_N (t) } \over {dt} } = 
   -N (B ( 1 - \delta_{N, N_{max}} ) + D) p_N (t) +
$$
\be
   (N+1) D (1 - \delta_{N, N_{max}}) p_{N+1}(t) + 
   (N-1) B (1 - \delta_{N, 0}) p_{N-1}(t)  
\quad ,
\label{eq3}
\ee
where $\delta_{ij}$ is $1$ if $i=j$ and zero otherwise. 
These equations are similar to ones appearing in 
population models (Nisbet \& Gurney, 1982).  
They have a simple physical interpretation: 
When $N$ species are present, 
any $N$ of them may go extinct with probability per unit time of $D$ 
or any $N$ of them may give rise to a new species with a probability 
per unit time of $B$. 
These two processes move the system out of the case of $N$ species 
and reduce $p_N$ by $N ( D + B)$ times the probability that the 
system has $N$ species, that is, $p_N$.  
The first term on the right hand side (RHS) 
of eq.\bref{eq3} 
represents these two processes. 
If the system has $N+1$ species, 
then any one of the $N+1$ species may go extinct with a probablity 
per unit time of $D$ to arrive in the $N$-species sector. 
This increased $p_N$ by $(N+1) D$ times the probability 
that the system has $N+1$ species. 
This generates the second term on the RHS 
of eq.\bref{eq3}. 
Finally, 
if the system has $N-1$ species, 
then any one of the $N-1$ species may give rise to a new species 
with a probablity 
per unit time of $B$ to put one in the $N$-species sector. 
This explains the third term on the RHS 
of eq.\bref{eq3}. 
The delta functions 
in  eq.\bref{eq3} 
take into account the ``boundary condition'' effects
when $N=0$ and when $N = N_{max}$.
 
We initially tried to find solutions to eq.\bref{eq3} 
using the diffusion approximation, 
which is the same approach used in Foley (1997) for populations. 
However, the results, when compared to simulated output, 
were not very accurate: 
The diffusion approximation works best when $N$ 
is large so that $N/N_{max}$ can be treated 
as a continuous variable.
In our case, $N_{max}$ is not sufficiently big. 

It follows from multiplying eq.\bref{eq3} by $N$ 
and summing over $N$ that 
the average number of species $\bar N(t)$ obeys 
$
 { {d \bar N(t)} / {dt}} = (B-D) \bar N(t) - N_{max} B p_{N_{max}} (t)
$. 
This equation implies exponential growth in time for $\bar N(t)$ 
with a coefficient $G$
during Stages (1) and (2), 
because the constraint that $N (t) \le N_{max}$ 
can be neglected. 

Assembling $p_N (t)$ for $N=0, 1, . . . , N_{max}$ 
into an $N+1$ column vector $\Psi$, 
allows the dynamical system 
governed by eq.\bref{eq3} to be cast in the form 
\be
  { {d \Psi} \over {dt} } = - H \Psi
\quad .
\label{eq4}
\ee
To solve eq.\bref{eq4},  
one needs to find the right-side eigenvalues $E_i$ 
and eigenvectors $v^{(i)}$ 
of $H$. 
Then the exact solution 
is $p_N(t)= \sum_{i=0}^{N_{max}} c_{(i)} v^{(i)}_N exp(-E_i t)$,  
where the coefficients $c_{(i)}$ are determined  
by the initial conditions  
$p_N(0)= \sum_{i=0}^{N_{max}} c_{(i)} v^{(i)}_N$.  
Since $E_i > 0$ for $i >0$,  
one sees that, as $t \to \infty$,   
$p_N( t) \to \delta_{N 0} = v_N^{(0)}$ and 
that $c_{(0)}$ must be $1$.  
This proves that a genus eventually goes extinct  
for finite $N_{max}$.  

It turns out that   
all eigenvalues are positive except for one,  
$E_0$, which is zero and corresponds to the situation  
of an extinct genus with $p_0(t)=1$.  
The smallest non-zero eigenvalue determines  
the asymptotic dynamics of the evosystem  
and the behavior of Stage (3).  
We have found that as soon as $N_{max}$ is sizeable  
(greater than say 6),  
there is a single, very small eigenvalue $E_1$. 

The genus goes extinct
at an average time scale of $T_e = 1/ E_1$, 
which is much larger than any of the time parameters 
of the evosystem.  
We have uncovered an accurate approximate formula for $E_1$:  
\be
 E_1 \approx { {N_{max} ! D^{N_{max}} } 
   \over { \sum_{j=0}^{N_{max}-1} 
   a_j^{(N_{max})} B^j D^{N_{max}-j-1} } } 
\quad .
\label{eq5}
\ee
where the coefficients $a_j^{(N_{max})}$ are determined recursively: 
$a_j^{(N_{max} )} = N_{max} a_j^{(N_{max} - 1 )} + (N_{max}-1)!$ 
for $j=1, 2, \dots N_{max}-2$, 
and $a_{N_{max}-1}^{(N_{max})}=(N_{max}-1)!$. 

For comparison, when $N_{max} =10$, and $B/D=2$, 
$E_1/D=0.0430\dots$ is the exact result, 
whereas eq.\bref{eq5} gives $E_1/D \approx 0.0427$. 

We have found an accurate formula for $v^{(1)}$:   
Set $v^{(1)}_{N_{max}}=1$ and 
$v^{(1)}_{N_{max-1} } = (N_{max} D - E_1)/((N_{max} -1) B)$ 
and then use 
$
 (N-1) B v^{(1)}_{N-1} = 
   (N (B+D) - E_1) v^{(1)}_N - (N+1) D v^{(1)}_{N+1}
$, 
for $N=N_{max}-1$ to $2$ 
and finally set 
$v^{(1)}_{0} = - \sum_{N=1}^{N_{max}} v^{(1)}_{N}$. 
Since $v^{(0)}$ is also known,  
an approximate analytic solution  
for the $p_N (t)$ during Stage (3)  
is obtained by using these lowest two eigenstates of $H$:  
\be
  p_N (t) \approx v_N ^{(0)} + 
    ((D/B)^{N_0} -1) v_N ^{(1)}/v_0^{(1)} \exp [-E_1 (t-t_{asy})] 
\quad , 
\label{eq6}
\ee
where $t_{asy} = 7/(4 (B-D) ) ln[N_{max}/N_0]$ is 
the time that it takes to reach Stage (3) on average. 
For example, 
in dimensionless units of time, 
let $D=0.1$, $B=0.2$, and $N_{max}=10$. 
Then at $t=60$, 
the approximate probabilities $(p_0, p_1, \dots p_{10})$ 
of eq.\bref{eq6} are 
(0.513, 0.0021, 0.0031, 0.0049, 0.0078, 0.0129, 0.0219, 
0.0378, 0.0663, 0.118, 0.213) 
and can be compared to the exact results of 
(0.504, 0.0025, 0,0036, 0.0053, 0.0083, 0.0134, 0.0224, 
0.0385, 0.0673, 0.120, 0.215). 
This excellent agreement is typical. 
The derivation of the approximate solution is outlined in Appendix B. 

For $N_{max}=\infty$, 
the solution to eq.\bref{eq6} is 
known (Nisbet \& Gurney, 1982):  
$$
 p_0(t) = {{{D} \left(  {e^{G t} - 1} \right) }\over 
   { B {e^{G t}} - D }} 
\quad , 
$$
\be
   p_N(t) = {{{B^{N-1}} {{\left( B - D \right) }^2}  
      {e^{G t}} {{\left( {e^{G t}} - 1 \right) }^{N-1}}}\over 
    {{{\left( B {e^{G t} - D} \right) }^{N+1}}}}
\quad {\rm for \ } N > 1 \quad . 
\label{eq7}
\ee
Note that $p_0(t)$ approaches eq.\bref{eq2} for $N_0 = 1$ 
when $t$ is greater than $1/G$ 
as should be the case. 
When $N_0 > 1$, 
$p_0(t)$ is given as in eq.\bref{eq7} above but with the RHS  
raised to the $N_0$th power. 
Because the $N_{max}$ constraint plays no role during Stage (1), 
eq.\bref{eq7} accurately approximates the $p_N (t)$ for the finite 
$N_{max}$ case at early times for the cases where 
$N$ is somewhat less than $N_{max}$. 

\medskip

{\bf\large\noindent V.\ Punctuated Equilibrium}
\vglue 0.2cm

Based on the fossils of the Burgess Shale, 
a new picture of diversity
in the tree of life has been suggested  
(Gould et al., 1977; Valentine, 1969; Sepkoski, 1978)
that involves initial growth and 
subsequent decimation and restriction 
(see for, example Figure 1.17 in Gould (1989)   
and see the discussion in Valentine \& Erwin (1985)). 
To check the validity of this idea, 
we incorporate catastrophes in the computer simulations 
by selecting a point in time 
and randomly rendering a species extinct with probability $p_c$. 
Figure 2b displays a typical output. 
The catastrophic event usually resets the genus system to Stage (1). 
If the genus survives the period of precariousness, 
exponential expansion occurs 
re-filling the available phase space of the evosystem. 
The phylogenetic trees outputted in our simulations 
do not resemble the new picture 
because the effects of decimation due to a catastrophe 
are eliminated exponentially quickly on a time scale of $t_{asy}$, 
which is of order $1/(B-D)$. 
The rapid radiation that occurs after a mass extinction 
is well documented (Sepkoski, 1993; Jablonksi, 1995). 
See Erwin et al. (1988) 
for a comparison of diversity patterns 
during the Paleozoic and Mesozoic eras. 

The picture of the diversity of life should be as follows: 
A significant addition to the span of diversity occurs 
whenever new phase space in the evosystem becomes available. 
See Figure 2c. 
Catastrophes should slightly decrease diversity 
but not drastically because a subset of species 
is able to maintain most of the morphological features 
of the larger set. 
During the rest of the time, 
when genera are in Stage (3), 
the change in diversity should be gradual: 
Adaptation can lead to new developments that 
slowly increase diversity, 
while convergent evolution 
or the haphazard extinction of a particular species 
may reduce it. 

The typical phylogenetic branch is obtained by 
piecing together Figures 1a, 2b and 2c in various ways. 
This picture is confirmed by the fossil record. 
See, for example, MacFadden \& Hulbert (1986). 

Our approach can quantify the {\it importance} 
of punctuated 
equilibrium (Eldredge \& Gould, 1972; Stanley, 1979; Gould \& Eldredge, 1993), 
an issue that has been hotly debated. 
Much evidence has been gathered that supports the role of 
punctuated equilibrium in evolution 
(Cheetham, 1986; Stanley \& Yang, 1987; Jackson \& Cheetham, 1990). 
If $r_c$ catastrophic extinction events occur per unit time, 
then the rate of speciation for ${\cal G}$ 
due to catastrophes is roughly $r_c p_c N_{max}$. 
During ``equilibrium periods,'' 
the rate of species growth is $D N_{max}$, as noted above. 
The ratio $r_c p_c / D$ of these two rates provides 
the relative importance of speciation for the two situations. 
For $r_c \approx $ one event per 40 million years, 
$p_c \approx 3/4$, and 
a genus with $D =$ one species per 300,000 years, 
this ratio is only 1\% so that 
speciation during equilibrium periods is much more important 
than speciation just after a catastrophe. 
Punctuated periods also occur 
when a new region of evosystem phase space becomes available. 
During this radiation, a genus starts in Stage (2), 
increases its number of species 
from $N_{max}^{old}$ to $N_{max}^{new}$ via exponential growth, 
and then arrives at Stage (3). 
See Figure 2c. 

Assuming that species generation proceeds slowly 
during stable geological periods, 
one concludes that punctuated equilibrium is relatively unimportant 
for speciation. 
However, 
visually punctuated events are quite striking as Figure 2 shows. 
They also often mark a significant change 
in the direction of evolution. 
Of course, 
proponents of punctuated equilibrium might argue 
that new species arise rapidly even during geologically calm periods. 
On the other hand, 
opponents of punctuated equilibrium might argue 
that speciation might be occuring gradually even just after catastrophes. 
Our mathematical analysis cannot address either of these issues.  

\medskip

{\bf\large\noindent VI.\ An Application of the Analytic Results}
\vglue 0.2cm

Our methods can help to address the controversy 
over the survival of Cambrian phyla. 
The issue is whether unusual phyla of the Cambrian Period 
have survived to the present. 
S.\,J.\,Gould (Gould, 1989) argues that many phyla 
are unique to the Cambrian Period,  
while S.\,Conway Morris (Morris, 1998) takes the opposite view 
that descendents of most Cambrian phyla have survived 
to the Recent.  

Stage (1) precariousness tells us that 
$D/B$ of Cambrian phyla consisting of a single species 
should have gone extinct in a relatively short period of time. 
Hence, a small but reasonable fraction of the Cambrian phyla 
should not be associated with lineages surviving to the present. 
If the evosystem did not allow more than about 10 members 
for a particular phylum, 
then that phylum probably did not survive to the Recent 
due to Stage (3) contingency. 
For example, if $B=$1/(100,000 years), 
$D=$1/(200,000 years), and $N_{max}=10$,  
then $T_e \approx$ 90 million years  
and it is very unlikely that such a phylum survived 
beyond the Paleozoic Era. 
However, if $B=$1/(500,000 years), 
$D=$1/(1 million years), and $N_{max}=10$, 
$T_e \approx$ 470 million years and 
roughly $1/3$ of such phyla that achieved Stage (3) status 
would survive to the present. 
This analysis, however, 
does not include the effects of mass extinctions, 
which decrease the chances of survival of lineages. 

\medskip

{\bf\large\noindent VII.\ Two-Genera Systems}
\vglue 0.2cm

In the real world, a genus does not evolve in isolation 
since organisms compete for shared resources. 
This introduces phylogenetic ``interactions''. 
The simplest way to incorporate these effects 
is to consider two genera ${\cal G}_a$ and ${\cal G}_b$ 
that share the same ecosystem 
with a constraint of $N_{max}$ on the combined number of species: 
$N_a + N_b \le N_{max}$. 
Since the birth and death rates 
for ${\cal G}_a$ and ${\cal G}_b$ may differ, 
there are four additional parameters: 
$B_a$, $D_a$, $B_b$ and $D_b$, 
where subscripts ``a'' and ``b'' distinguish the two genera. 

As is the case of the one-genus system, 
there are usually three stages of evolution. 
If both genera start with a small number of species, 
then Stage (1) is characterized by precariousness: 
Either genera may go extinct by ``bad luck.'' 
If this happens to only one of the two genera, 
the two-genus system becomes a one-genus system and 
the results above apply. 
Assuming that both genera survive Stage (1), 
the genera enter Stage (2), 
which is characterized by exponential growth: 
$N_a \sim \exp (G_a t)$ and $N_b \sim \exp (G_b t)$ 
with $G_a = B_a - D_a$ and $G_b = B_b - D_b$. 
Both genera expand rapidly 
until the effect of the constraint 
on the maximum number of species is felt. 
At this point, 
Stage (3) arrives and the two genera compete 
for limited evosystem phase space. 

During Stage (3), 
anything is possible since 
a stochastic process is involved: 
Either ${\cal G}_a$ or ${\cal G}_b$ may go extinct first. 
To address what happens on average, 
introduce $p_{N_a , N_b} (t)$,  
which are the probabilities that the evosystem has 
$N_a$ species of ${\cal G}_a$ and $N_b$ species of ${\cal G}_b$ 
at time $t$. 
The differential equations that determine 
the time evolution of $p_{N_a , N_b} (t)$ 
are similar to those to eq.\bref{eq3} 
and straightforward to write down 
but are more complicated due to the presence of two genera 
and the boundary conditions of 
$N_a \ge 0$, $N_b \ge 0$ and $N_a + N_b \le N_{max}$ that 
determine the evosystem phase space. 
Assembling $p_{N_a , N_b}$  
into an $(N_{max}+1) (N_{max}+2)/2$ column vector $\Psi$,  
the system of equations can be again written as in eq.\bref{eq4},  
and because it is linear,  
it is exactly solvable once the eigenvalues and eigenvectors  
are found. 

The real parts of all the eigenvalues are positive except for one,  
which is zero and corresponds to both genera being extinct.  
The eigenvalues of the single-genus system  
with parameters $B_a$, $D_a$, and $N_{max}$  
are also eigenvalues of the two-genus system. 
The reason is as follows: 
if $v_N^{(i)}$ is an eigenvector for the single-genus system 
then $v_{N_1,N_2}^{(i)} = v_{N_1}^{(i)} \delta_{N_2 , 0}$ 
and $v_{N_1,N_2}^{(i)} = \delta_{N_1 , 0}v_{N_2}^{(i)} $ 
are eigenvectors for the two-genus case.    
It turns out that there are only two very small eigenvalues,  
and they are equal to those of the single-genus systems,  
for which eq.\bref{eq5} provides accurate approximations. 

Let $E_{1a}$ (respectively, $E_{1b}$) denote the smallest 
non-zero eigenvalue of $H$ 
for the one-genus with parameters $B_a$, $D_a$ and $N_{max}$ 
(respectively, $B_b$, $D_b$ and $N_{max}$).   
If both genera arrive at Stage (3), 
then usually the genus with the smallest $E_{1}$ 
eventually dominates 
with the other genus going extinct first. 
Therefore quite often, 
the genus with the smallest death to birth rate $D/B$ 
survives the longest. 
Species with ``conservative'' habits such as burrowing organisms, 
nocturnal creatures, and life with strong protective features, 
are more likely to have smaller $D/B$ ratios 
because they are less susceptible to destructive natural forces 
such as predation and geological calamities. 
Therefore, such species often have a better chance to survive for long times. 
Among such species, 
some may also have a smaller birth rate $B$. 
This can lead to a ``tortoise and hare'' effect: 
Imagine a situation for which $G_a$ is bigger than $G_b$ 
but with $E_{1b}$ less than $E_{1a}$. 
Then, on average, 
genus ${\cal G}_a$ (``the hares'') more quickly radiate initially 
but genus ${\cal G}_b$ (``the tortoises'') 
usually eventually dominate. 
The computer simulation in Figure 3a illustrates the effect. 

It has been known for a long time that 
populations of prey and predator 
can undergo oscillations (Utida, 1957; Leslie \& Grower, 1958).
The question is 
whether similar oscillatory effects arise 
among genera of species of prey and predators. 
Let the letter ``a'' denote a predator genus and 
let ``b'' denote a prey genus. 
Assume that the ecosystem can support 
up to $N_{max}^a$ predator species and $N_{max}^b$ prey species, 
and 
let the birth and death rates of ${\cal G}_a$ and ${\cal G}_b$ 
depend on $N_a$ and $N_b$. 
We choose 
$D_b = D_b^0 + D_{ba} N_a$ with $D_{ba}$ being a constant,  
$B_a = B_{ab} N_b$ with $B_{ab}$ being a constant, 
and we take the parameters $D_a$ and $B_b$ 
to be independent of $N_a$ and $N_b$. 
The functional forms for $D_b$ and $B_a$ 
are reasonable if the number of prey (respectively predator) individuals 
is proportional to the number of prey (respectively predator) species. 
This is more likely than not to be the case, 
but there may be specific examples when it is not true. 

One arrives at the Volterra-Hamiltonian models (Hastings, 1997) 
by using a ``mean field theory'' approximation 
for which the stochastic variables $N_a$ and $N_b$ 
are replaced by the mean values $\bar N_a$ and $\bar N_b$. 
During Stage (2), 
the growth rates of $G_a$ of ${\cal G}_a$ and $G_b$ of ${\cal G}_b$ 
are respectively 
$G_a = B_a - D_a \approx B_{ab} \bar N_b - D_a$
and
$G_b = B_b - D_b \approx B_b - D_b^0 - D_{ba} \bar N_a$. 
A fixed point exists at $\bar N_b = D_a/B_{ab}$ and 
$\bar N_a = (B_b - D_b^0)/D_{ba}$. 
With appropriate initial conditions, 
this system often exhibits oscillations. 
Figure 3b shows one of our computer simulations 
that generates such behavior. 
The current fossil record is undoubtedly too fragmented 
to see oscillations in species numbers 
but, because of the vast richness of ecosystems and life forms
during the long history of the Earth,  
it is likely that there were situations in evolution 
for which our prey-predator model is sufficiently good and 
for which such oscillatory effects did occur. 

Interestingly, 
we found that only a limited number of oscillations took place. 
To determine what happens on average, 
let $p_{N_a, N_b} (t)$ be the probability 
that there are $N_a$ members of ${\cal G}_a$ and 
$N_b$ members of ${\cal G}_b$ at time $t$. 
Then the system of equations 
for the $p_{N_a, N_b}$ can again be cast as 
$d \Psi / dt = - H \Psi$. 
Some of the eigenvalues of $H$ appear in complex pairs, 
signaling the possibility of oscillatory behavior. 
However, the real parts of the complex eigenvalues are sizeable 
compared to the imaginary parts, 
thereby causing oscillations to damp out. 
Figure 3c shows the behavior of $\bar N_a (t)$ and $\bar N_b (t)$ 
for a typical case. 
We arrive at the perhaps surprising conclusion that, 
while specific evolutions may exhibit many oscillations, 
{\it on average this does not happen}.

\medskip

{\bf\large\noindent VII.\ Summary}
\vglue 0.2cm

Our computer simulations reveal that a branch of the evolutionary tree 
corresponding to a genus has three states: 
During Stage (1), 
there is a sizeable chance that the genus goes extinct, 
Figure 1b being an example. 
The probability that this happens 
is given in eq.\bref{eq2} 
where $N_0$, 
the initial number of species in the genus, 
is usually one. 
Stage (2) is characterized by exponential growth in time 
with $1/(B-D)$ being the time constant. 
In Stage (3), 
the number of species lingers near $N_{max}$. 
However, this stage is quasi-stable in the sense 
that extinction  
will eventually happen but at a long time 
scale that is equal to $1/E_1$ on average where 
$E_1$ is given in eq.\bref{eq5}. 
Any genus of species is guaranteed to perish at some point. 

There are differential equations (eq.\bref{eq3}) 
that determine the probability $p_N(t)$ that a genus 
has $N$ species at time $t$. 
For Stage (1), 
the $p_N (t)$ are well approximated by the $N_{max} = \infty$ 
case, 
for which an exact result is provided 
in eq.\bref{eq7}. 
For Stage (3), 
we obtained an accurate but approximate solution 
(see eq.\bref{eq6}). 

We have identified two new types of contingency: 
(a) extinction associated with ``bad luck'' during Stage (1) 
and 
(b) extinction associated with the quasi-stability of Stage (3). 
Note that although a genus will disappear on average 
at a time given by $1/E_1$, 
there is great variation in this extinction time for specific cases 
(that is, one genus may perish at $5/E_1$ while other may only 
survive as long as $0.1/E_1$). 

We have shown that due to Stage (1) contingency, 
the extinction of a genus during a catastrophic event 
may not be immediate but may occur over a time scale set by $1/B$. 
Figure 2a is an example from a computer simulation. 
Thus, at the genus level, 
a ``smearing in time'' arises for catastrophic mass extinctions. 

Our analytic and simulation methods produce a new picture of diversity: 
Diversity should initially increase exponentially and then slowly 
increase with small setbacks due to mass extinctions. 
This picture combines features of gradualism and punctuated equilibrium. 

Using reasonable values for the time scales of speciation and catastrophes, 
we show that speciation during ``calm'' periods is more important 
than just after a catastrophe. 
If speciation during a catastrophic rebound is 100\% attributable 
to punctuated equilibrium 
and speciation during other times is gradual 
then the generation of new species from punctuated periods  
is roughly two orders of magnitude smaller than during other times. 
  
There should be situations in Earth's evolutionary past 
exhibiting the ``tortoise and the hare'' effect. 
One genus (the hares) initially increase its number of species rapidly 
while the other genus (the tortoises) lag behind, 
but eventually the slower genus (the tortoises) survive 
longer than the faster genus (the hares) because the former 
has a small species death rate $D$. 
Figure 3a illustrates this. 
We also argue that, in certain, fairly special situations, 
oscillations in the number of species of a genus may occur. 
See Figure 3b. 
On average, 
the oscillatory behavior is damped 
so that only a few oscillations occur. 

Complex behavior such as chaos cannot arise for averaged quantities: 
Since all the above systems governed by $H$ are linear and 
exactly solvable, they cannot be chaotic, 
although the evolution of a particular genus 
is stochastic and involves considerable random effects. 

\medskip

{\bf\large\noindent Appendix A: A Derivation 
of the Extinction Probability that Arises During Stage (1)}
\vglue 0.2cm

Eq.\bref{eq2} for $N_0 =1$ 
can be derived using using the idea of self-similarity. 
Let $\Delta t$ be a small time interval and set 
$d = D \Delta t$, $b = B \Delta t$ and $u = 1 - b - d$. 
A single line may survive $k$ time intervals  
and then go extinct (which happens with probability $u^k d$), 
or it may survive $k$ time intervals and then split into two  
(which happens with probability $u^k g$). 
Summing the single line graph cases gives $d/(1-u)$ 
since $k$ may range from $0$ to $\infty$. 
When a single line splits into two,
two copies of the original system are obtained.   
The probability that both lines go extinct is $P_e^2$. 
Summing the graphs for the splitting case gives $P_e^2 g/(1-u)$. 
Combining, one obtains $P_e = d/(1-u) + P_e^2 g/(1-u)$. 
Solving this quadratic equation for $P_e$ 
generates the result for $P_e$ given in Section III. 

\medskip

{\bf\large\noindent Appendix B: A Derivation 
of the Approximate Asymptotic Solution}
\vglue 0.2cm

The late time behavior of the $p_N (t)$ is governed by 
the eigenvalues $E_i$ of $H$ 
in eq.\bref{eq4} 
with the smallest real parts. 
We numerically calculated the right-side eigenvalues of $H$
and discovered that there are only two relevant ones 
no matter what are the values of $B$, $D$ and $N_{max}$. 
For one-genus systems, 
all eigenvalues are real and non-negative (the latter must be 
true if probability is to be conserved). 
One eigenvalue $E_0 = 0$ corresponds to extinction; 
its eigenvector $v^{(0)}$ is 
$(1, 0, 0, \dots , 0)$ and corresponds to $p_N = \delta_{0 N}$. 
The other small eigenvalue $E_1$, 
which is given 
in eq.\bref{eq5}, 
was obtained by computing the characteristic equation 
$det (H - I \lambda) = 0$ 
and neglecting terms of order $\lambda^2$ and higher. 
This linearization is justified because $\lambda$ is small. 
Even after this simplification, 
it is not easy to find the solution $\lambda$ ($\equiv E_1$) 
because the determinants can be sizeable and results 
are fairly complicated functions of $B$, $D$ and $N_{max}$. 
We proceeded by obtaining the solutions for $N_{max} = 2$ through $6$. 
We noticed patterns in various coefficients that led to the recursion 
relations for $a_j^{ ( N_{max} )}$ given 
below eq.\bref{eq5}. 
We then checked ``our guess'' for $N_{max} = 7$ and $8$
to confirm our formula. 

To obtain the approximate eigenvalue $v^{(1)}$, 
use the result for $E_1$ 
in eqs.\bref{eq3} and \bref{eq4} 
and solve the equation set for the $p_N$ starting with $N= N_{max}$ and 
working down to $N = 1$. 
One finds the result for $v^{(1)}$ given above eq.\bref{eq6}.  
Saturating the general solution with the two smallest eigenvalues 
produces the form of the general solution: 
\be
  p_N(t) \approx c_{(0)} v^{(0)}_{N} 
    \exp ( -E_0 t) + c_{(1)} v^{(1)}_{N} exp(-E_1 t)
\quad  . 
\label{eq8}
\ee 
It remains to determine $c_{(0)}$ and $ c_{(1)}$. 
As explained in Sect.\,IV, 
$c_{(0)}=1$. 
Let $t_{asy}$ be the time at which Stage (3) is established. 
Beyond this time,  
$p_0 \approx (D/B)^{N_0}$. 
Substituting this result into the approximate solution, 
one finds
$$ 
 c_{(1)} \approx 
  \left( { \left( { {D \over B} } \right)^{N_0} -1 } \right) 
  { { \exp ( E_1 t_{asy} ) }  \over {v^{(1)}_{0}} }
\quad .
$$
To determine 
$t_{asy}$, 
note that it is of order $1 / (B-D) ln \lbrack N_{max} / N_0 \rbrack $ 
due to the exponential growth behavior of Stage (2). 
We found by comparison with numerical solutions that 
$t_{asy} = 7/4 ln \lbrack N_{max} / N_0 \rbrack / (B-D)$ produced 
good results. 
Although $7/4$ might not be precisely the optimal choice
for the coefficient, 
agreement at about the 1\% occurs with this value.

\medskip 
{\bf\large\noindent Acknowledgments}  

We thank Ravi Iyengar for assistance.  
This work was supported in part 
by the National Science Foundation under the grants PHY-9420615 
and PHY-0098840, 
by the Aaron Diamond Foundation, 
and by the Director, Office of Science, Office of High Energy 
and Nuclear Physics, of the U.\,S.\,Department of Energy 
under contract DE-AC03-76SF00098. 

\medskip 

\begin{center} 
{\bf\large REFERENCES} 
\end{center} 

Alvarez, L., Alvarez, W., Asaro, F.\,\& Michel, H.\,V. (1980), 
``Extraterrestrial Cause for the Cretaceous-Tertiary Extinction'',
Science {\bf 208}, 1095-1108.

Archibald, J.\,D. (1981),
``The Earliest Known Palaeocene Mammal Fauna and Its Implication 
for the Cretaceous-Tertiary Transition'', 
Nature {\bf 291}, 650-652.

Cheetham, A.\,H. (1986), 
``Tempo of Evolution in a Neogene Bryozoan: 
Rates of Morphologic Change Within and Across
Species Boundaries'',
Paleobiology {\bf 12}, 190-202.

Conway Morris, S. (1998),  
{\it The Crucible of Creation: The Burgess Shale 
and the Rise of Animals}   
(Oxford University Press, Oxford). 

Eldredge, N.\,\& Gould, S.\,J. (1972),  
in {\it Models in Paleobiology},  
ed.\, T.\,J.\,M.\,Schopf,  
(Freeman Cooper, San Francisco),  
82-115. 

Erwin, D., Valentine, J., and Sepkoski, J.\ (1988)
``A Comparative Study of Diversification Events'', 
Evolution {\bf 41}, 1177-1186.

Foley, P.\ (1997), 
``Extinction Models for Local Populations'' 
in {\it Metapopulation Biolgy}, 
ed. Hanski, I.\, \& Gilpin, M.\,E., 215-245,
(Academic Press) 
and references therein. 

Gould, S.\,J., Raup, D.\,M., Sepkoski, J.\,J., 
Schopf, T.\,J.\,M.\,\& Simberloff, D.\,S. (1977),  
``The Shape of Evolution: A Comparison of Real and Random Clades'',  
Paleobiology {\bf 3}, 23-40. 

Gould, S.\,J. (1989),  
{\it Wonderful Life} (W. W. Norton, New York). 

Gould, S.\,J.\,\& Eldredge, N. (1993),   
``Punctuated Equilibrium Comes to Age'',  
Nature {\bf 366}, 223-227.  

See for example Hastings, A. (1997),  
{\it Population Biology} (Springer Verlag, New York). 

Hickey, L.\,J. (1981), 
``Land Plant Evidence Compatible with Gradual, 
Not Catastrophic Change
at the End of the Cretaceous'',
Nature {\bf 292}, 529-531.

Jablonksi, D. (1995), 
``Extinctions in the Fossil Record'', 
in {\it Extinction Rates}, 
ed. Lawton, J.\,\& May, R.\,M., 1-21, 
(Oxford University Press) and references therein.

Jackson, J.\,B.\,C.\ and A.\,H.\,Cheetham (1990),
``Evolutionary Significance of Morphospecies: 
a Test with Cheilostrome Bryozoa'',
Science {\bf 248}, 579-583.

Leslie, P.\,H.\,\& Grower, J.\,C. (1958), 
``The Properties of a Stochastic Model 
for the Predator-Prey Types of Interactions Between Two Species'', 
Biometrika {\bf 57}, 219-234. 

MacArthur, R.\,H. (1972), 
{\it Geographical Ecology}
(Harper \& Row); see the Appendix on page 121. 

MacFadden, B.\,J.\,\& Hulbert, Jr.,R.\,C. (1986),
``Explosive Speciation at the Base 
of the Adaptive Radiation of Miocene Grazing Horses'', 
Nature {\bf 336}, 466-468.                              

May, R.\,M., Lawton, J.\,\& Stack, N.\,E.\ (1995), 
in {\it Extinction Rates}, ed. Lawton, J.\,\& May, R.\,M., 
(Oxford University Press) 1-21. 

McCune, A.\,R. (1997), 
``How Fast is Speciation'' 
in {\it Molecular Evolution and Adaptive Radiation}, 
ed. by Gvnish T.\,J.,  and Sytsma, K.\,J.,
(Cambridge University Press). 

Nisbet, R.\,M.\,\& Gurney, W.\,S.\,C. (1982),
{\it Modelling Fluctuating Populations}, 
see Chapter 6,
(John Wiley \& Sons). 

Renshaw, E. (1991), 
{\it Modelling Biological Populations in Space and Time} 
(Cambridge University Press, Cambridge).  

Sepkoski, J.\,J. (1993), 
``Ten Years in the Library: 
New Data Confirms Paleontological Patterns'', 
Paleobiology {\bf 19}, 43-51. 

Sloan, R.\,E., Rigby Jr., J.\,K., Van Valen L.\,M.\,\& Gabriel, D. (1986), 
``Gradual Dinosaur Extinction and Simultaneous Ungulate Radiation 
in the Hell Creek Formation'',
Science {\bf 232}, 629-33.

Sepkoski, J.\,J. (1978), 
``A Kinetic Model of Phanerozoic Taxonomic Diversity I: 
Analysis of Marine Orders, 
Paleobiology {\bf 4}, 223-251. 

Stanley, S.\,M. (1979),
{\it Macroevolution: Patterns and Process} 
(W.\,H.\ Freeman).

Stanley, S.\,M., and Yang, X. (1987), 
``Approximate Evolutionary Stasis 
for Bivalve Morphology over Millions of Years: 
a Multivariate, Multilineage Study'', 
Paleobiology {\bf 13}, 113-139.

Utida, S. (1957), 
``Cyclic Fluctuations of Population Density Intrinsic 
to the Host-parasite System'', 
Ecology {\bf 38}, 442-429. 

Valentine, J. (1969), 
``Patterns of Taxonomic and Ecological Structure 
of the Shelf Benthos During Phanerozoic Time'',
Paleaontology {\bf 12}, 684-709.

Valentine, J., and Erwin, D.\ (1985)
``Interpreting Great Developmental Experiments: The Fossil Record'',
in {\it Development as an Evolutionary Process}
ed.\, Raff, R.,\,A.\ and Raff, E.\,C.,
(Alan R. Liss, Inc., New York) 71-107.


\medskip 
{\bf\large\noindent Figure Captions}  

\noindent
{\bf Figure 1} Single Species Simulations with Parameters 
$b=0.25$, $d=0.1$, $u=0.65$ and $N_{max} = 15$. \\ 
{\bf a} The Development of a Genus Showing the Three Stages. \\ 
{\bf b} A Genus that Does Not Survive Stage (1). 

\noindent
{\bf Figure 2} Simulations with Punctuated Equilibrium with 
Parameters $b=0.25$, $d=0.1$, and $u=0.65$. \\
{\bf a} ``Slow Extinction'' after a Catastrophe.   
A catastrophic event with $p_c=0.9$ occurs as indicated, 
but the genus does not survive Stage (1) and goes extinct 
after a relatively long time period. 
$N_{max}$ is $15$. \\
{\bf b} The Usual Effects of a Catastrophe. 
A catastrophic event with $p_c=0.9$ occurs as indicated, 
and the genus survives Stage (1) eventually filling 
the evosystem phase space. 
The simulation was a continuation of that in Figure 2a. \\ 
{\bf c} The Effects of New Phase Space. \\
At $t_{exp}$, $N_{max}$ goes from $10$ to $20$ 
leading to a radiation.

\noindent
{\bf Figure 3} Simulations with Two Genera. \\
{\bf a} The ``Tortoise and the Hare'' Effect. 
${\cal G}_a$, the ``aggressive'' genus, initially dominates but, 
on the long run, ${\cal G}_b$, the ``conservative'' genus, survives 
while genus ${\cal G}_a$ goes extinct. 
The parameters for the simulation are $N_{max} =15$, 
$b_a =0.45$, $d_a =0.15$, $u_a =0.40$, 
$b_b =0.20$, $d_b =0.05$, $u_b =0.75$. 
For clarity, the evolution between times $20$ and $180$ 
is not shown. \\
{\bf b} Predator/Prey Oscillations. 
The number of predators and prey species oscillate. 
The curve for the predators leads the curve for the prey. 
The genus of prey goes extinct around $t=230$, 
and the extinction of the genus of predators soon follows. 
The parameters are $N_{max}^a = N_{max}^b =40$, 
$d_a =0.1$, $b_{ab} = d_{ba} =0.00625$, 
$b_b =0.101$, and $d_b^0 =0.001$. 
The lower case ``$b$'' and ``$d$'' indicate discrete versions 
of the continuous birth and 
death rate parameters ``$B$'' and ``$D$''. \\
{\bf c} Damped Predator/Prey Oscillations.  
The average number of predators $\bar  N_a$ and 
prey $\bar N_b$ undergo 
an initial oscillation, 
but eventually the they exhibits 
smooth behavior. 
The parameters are $N_{max}^a = N_{max}^b =40$, 
$D_a =2.0$, $B_{ab} = D_{ba} =0.25$, 
$B_b =2.5$, and $D_b^0 =0.5$.

\pagebreak

This document was prepared as an account of work sponsored 
by the United States Government. While this document is believed 
to contain correct information, neither the United States Government 
nor any agency thereof, nor The Regents of the University of California, 
nor any of their employees, makes any warranty, express or implied, 
or assumes any legal responsibility for the accuracy, completeness, 
or usefulness of any information, apparatus, product, or process disclosed, 
or represents that its use would not infringe privately owned rights. 
Reference herein to any specific commercial product, process, 
or service by its trade name, trademark, manufacturer, or otherwise, 
does not necessarily constitute or imply its endorsement, recommendation, 
or favoring by the United States Government or any agency thereof, 
or The Regents of the University of California. The views and opinions 
of authors expressed herein do not necessarily state or reflect those 
of the United States Government or any agency thereof, 
or The Regents of the University of California.

Ernest Orlando Lawrence Berkeley National Laboratory is an equal opportunity employer.

\vfill\eject 
\end{document}